\documentclass{mn2e}
\usepackage{epsf}
\begin{document}
%
% Title and front material
%
\title[Luminosity distance]{The Luminosity Distance in Perturbed FLRW
 Spacetimes} 
\author[Pyne \& Birkinshaw]{T. Pyne\\
        1101 East 58th Street, Chicago, IL 60615, USA\\
        \and M. Birkinshaw\\
        Department of Physics, University of Bristol, Tyndall Avenue,
        Bristol BS8~1TL}
\date{Submitted: 9-July-2003}
\maketitle
\label{firstpage}
\begin{abstract}
We derive an expression for the luminosity distance in FLRW
spacetimes affected by scalar perturbations. Our expression is
complete to linear order and is expressed entirely in terms of
standard cosmological parameters and observational quantities. We
illustrate the result by calculating the RMS scatter in the usual
luminosity distance in flat
$(\Omega_{\rm m},\Omega_\Lambda)=(1.0,0.0)$ and $(0.3, 0.7)$
cosmologies. In both cases the scatter is appreciable at high 
redshifts, and rises above 11\% at $z = 2$, where it may be the
dominant noise term in the Hubble diagram based on SN~Ia.
\end{abstract}

\section{Introduction}
\label{sec:intro}

A substantial body of modern theoretical cosmology is concerned with
the theory of small metric perturbations about
Friedman-Lema{\^\i}tre-Robertson-Walker (FLRW) cosmologies. In
particular, cosmologists often derive exact formulae for observables
in such spacetimes and then apply standard statistical techniques in
recognition of the stochastic nature of the perturbations. A great
benefit of such a programme is that correlations between effects
previously treated as distinct may be used to illuminate cosmological
questions.

As astronomers and cosmologists work with ever more distant
sources, it becomes increasingly important to understand and
account for the deviations of our Universe from the simple FLRW
spacetimes. One example, among many, where such an understanding
is important is in the use of type Ia supernovae as standard
candles to infer luminosity distance as a function of source redshift,
and hence to demonstrate that the Universe is currently in a phase of
accelerated expansion (Reiss et al.~1998; Perlmutter et al.~1999). The
inference that the Universe is 
accelerating was made on the basis of the luminosity distance formula
appropriate for the background cosmology, not to the perturbed
cosmology in which we live. The gravitational effects of large-scale
structure were taken to contribute to the error budget at a level
determined by the numerical studies of Wambsganss et al. (1997).
Gravitational lensing by large-scale structure is known to produce
effects on the order of those found in these numerical studies, and
this is usually assumed to be the dominant effect on the luminosity
distance (Kantowski, Vaughan \& Branch 1995; Holz \& Wald 1998; Holz
1998; Kaiser 1998). However, so far as we know, no rigorous
exploration of this assumption has been attempted. In any case, the
division of the effects of large-scale structure into categories such
as lensing, time-delays, and others is largely artificial: we measure
luminosities and redshifts in the real Universe and no idealized FLRW
background exists. 

In this paper we investigate the luminosity distance in linearly
perturbed FLRW spacetimes. Our results are obtained by direct
integration of the geodesic equations. A
complementary approach has been employed by Sasaki (1987). 
Our results improve on those of Sasaki in two ways. First, we present
explicit formulae for curved cases, whereas Sasaki's solutions in
these cases require a prior integration of the perturbed geodesic
equations. Second, our formulae are functions only of standard
cosmological parameters and observational quantities.

The plan of this paper is as follows. In section~\ref{sec:analysis} we
derive a formula for the luminosity distance, accurate to first order,
in metric perturbed FLRW spacetimes. In section~\ref{sec:lensing} we
evaluate the cosmological weak lensing correction to the usual
luminosity distance as a function of redshift in two model flat
spacetimes, with $\Omega_{\rm m}=1$ and $(\Omega_{\rm m},
\Omega_\Lambda)= (0.3, 0.7)$. Section~\ref{sec:summary}  
summarizes our results.

In this paper we use units such that $G = c = 1$. Greek indices
$\mu ,\nu ,...$ run over $\lbrace 0,1,2,3\rbrace $, and Roman indices
$i,j,...$ run over $\lbrace 1,2,3 \rbrace$. The spacetime metric
is taken to have signature $+2$, and the Riemann and Ricci tensor
conventions are given by $\left[ \nabla_{\alpha},\nabla_{\beta}\right]
v^{\mu}= R^{\mu}{}_{\nu\alpha\beta}v^{\nu}$ and $R_{\alpha\beta}=
R^{\mu}{}_{\alpha\mu\beta}$.

\section{Analysis}
\label{sec:analysis}

In any spacetime there is a relation between the element of
cross-sectional area of a radiation source, $dA$, and the solid
angle that this source is observed to subtend at an observer,
$d\Omega$. Although $dA$ is unobservable, this relation is usually
used to define a distance measure $r$, the angular-diameter distance,
according to $dA = r^2 \, d\Omega$. In a general spacetime $r$ will
depend upon the particular locations of the source and observer, and
possibly also on the choice of multiple null paths along which the
source is observed. In perfectly homogeneous and isotropic FLRW
spacetimes $r$ depends only upon the observed redshift of the source
and the cosmological parameters describing the evolution of the scale
factor: the density parameters in matter, curvature, and cosmological
constant and the Hubble constant, $\Omega_{\rm m}$, $\Omega_\kappa$,
$\Omega_\Lambda$ and $H_0$. In any spacetime, the luminosity
distance, $d_{\rm L}$, is related to $r$ and the observed redshift
$z$ of the source according to $d_{\rm L} = (1+z)^2 \, r$. In this
section we exploit our previous solution (Pyne \& Birkinshaw 1996)
for $r$ in linearly perturbed FLRW spacetimes to find $d_{\rm L}$.

We choose the cosmological model to be an FLRW spacetime with
embedded small scalar perturbations $\phi$ (the extension to
vector and tensor perturbations is conceptually straightforward).
The metric of such a spacetime may be written in the form
\begin{eqnarray}
  d{\bar s}^2 
    &=& a^2 \, \bigl[ -(1+2\phi ) d\eta^2 
        \nonumber \\
    & & + (1-2\phi )
          \gamma^{-2} \left( d{\hat x}^2 + d{\hat y}^2
          + d{\hat z}^2 \right) \bigr]
  \label{eq:metric}
\end{eqnarray}
where $\gamma = 1+ \kappa {\hat r}^2/4$, $\kappa$ is the spatial
curvature parameter ($\pm 1$ or 0), and ${\hat r}^2={\hat x}^2+{\hat
y}^2+{\hat z}^2$.\footnote{A hat is used merely to distinguish this
particular coordinate system, and not to denote a normalized vector.}

Consider the radiation emitted by a source at event ${\cal E}$,
corresponding to conformal time $\eta_{\rm e}$, that reaches an
observer at event ${\cal O}$ at conformal time $\eta_{\rm o}$ at
the spatial origin. Let the four-velocities of the source and observer
be written
$u_{\rm e}^{\mu}=\left( 1/a_{\rm e}\right)\left ( 1-\phi_{\rm e}
  ,v^i_{\rm e} \right)$
and
$u_{\rm o}^{\mu}=\left( 1/a_{\rm o}\right)\left ( 1-\phi_{\rm o}
  ,v^i_{\rm o} \right)$
respectively.\footnote{Note that these velocities are properly
normalized if their spatial components are treated as
first-order quantities.} Suppose, further, that the observer chooses a
spatial frame such that the source appears in the $\hat z$-direction
on the observer's sky. Then the relation between the cross-sectional
area of the source at ${\cal E}$, $dA$, and the solid angle the source
subtends at ${\cal O}$, $d\Omega$, was shown in Pyne \& Birkinshaw
(1996) to be
\begin{equation}
 dA = d\Omega \, \left( 1+2v^{\hat z}_{\rm o} \right) \,
      {a^2_{\rm e}{\hat r}^2_{\rm e} \over \gamma^2_{\rm e}}
      \, \left( 1-2\phi_{\rm e} \right) \, \det M
 \label{eq:sol}
\end{equation}
where
\begin{eqnarray}
  \det M &=& 1 \nonumber \\
         & & \hspace{-40pt} 
             + {4 \over \sin_\kappa \left( \eta_{\rm o}
             -\eta_{\rm
             e} \right)} \int_{\eta_{\rm o}}^{\eta_{\rm e}}d\eta \,
             \left( \eta -\eta_{\rm e}\right)\phi_{,\eta}
             \nonumber \\
         & & \hspace{-40pt} 
             -{4\over \sin_{\kappa}
             \left( \eta_{\rm o}-\eta_{\rm e}\right)}
             \int_{\eta_{\rm o}}^{\eta_{\rm e}}d\eta
             \, \sin_{\kappa}\left( \eta -\eta_{\rm e}\right)\gamma
             \phi_{,{\hat z}}
             \nonumber \\
         & & \hspace{-40pt} 
             +{2 \over \sin_{\kappa}
             \left( \eta_{\rm o}-\eta_{\rm e}\right)}
             \int_{\eta_{\rm o}}^{\eta_{\rm e}}d\eta
             \, \sin_{\kappa}\left( \eta -\eta_{\rm e}\right)
             \, {\hat r} \, \gamma \,
             \left[ \phi_{,{\hat x}{\hat x}}
                   +\phi_{,{\hat y}{\hat y}} \right]
  \label{eq:path}
\end{eqnarray}
the integrals being taken along the path $x^{(0)}=(\eta, 0,0,{\hat
r}(\eta)) $, which is a null geodesic of the background spacetime,
with affine parameter $\eta$, provided that ${\hat r}=2\tan_\kappa
((\eta_{\rm o} -\eta)/2)$. On this path $\gamma =\sec^2_\kappa
((\eta_{\rm o} -\eta)/2)$. In (\ref{eq:path}) and the following, the
subscript $\kappa$ on a trigonometric function denotes a set of
three functions: for $\kappa =1$ the trigonometric function
itself, for $\kappa =-1$ the corresponding hyperbolic function,
and for $\kappa =0$ the first term in the series expansion of the
function.\footnote{Pyne \& Birkinshaw (1996) write, more
precisely, $\lambda_{\rm e}$ where we write $\eta_{\rm e}$, but
the two quantities differ only at first order, which renders them
equivalent in (\ref{eq:sol}).}

The terms in (\ref{eq:sol}) have the following interpretations:
\begin{eqnarray*}
\left( 1+2v_{\rm o}^{\hat z}\right)
     &=& \pmatrix{\hbox{special relativistic transformation} \cr
                  \hbox{of solid angle at ${\cal O}$ due to} \cr
                  \hbox{observer's local peculiar velocity}
                 } \cr
     & & \qquad \null \\
{a^2_{\rm e} r^2_{\rm e} \over \gamma^2_{\rm e} }\left(
 1-2\phi_{\rm e} \right)
     &=& \pmatrix{\hbox{metric factors in induced} \cr
                  \hbox{area two-form in}          \cr
                  \hbox{rest frame of emitter}
                 } \cr
     & & \qquad \null \\
\det M
     &=& \pmatrix{\hbox{effect of the gravitational field}\cr
                  \hbox{on the light rays as they travel}\cr
                  \hbox{from source to observer} }
\end{eqnarray*}
In particular, the first two integrals in expression (\ref{eq:path}) for
$\det M$ describe longitudinal effects of the perturbations, while
the last term describes the transverse effects usually interpreted
as gravitational lensing.

It is convenient to write (\ref{eq:path}) in the form
\begin{eqnarray}
 \det M &=& 1 - 4\phi_{\rm o} \nonumber \\
        & & \hspace{-40pt}
            +{4 \over \sin_\kappa \left( \eta_{\rm
            o} -\eta_{\rm e}\right)} \int_{\eta_{\rm
            o}}^{\eta_{\rm e}}d\eta \, \left( \eta -\eta_{\rm
            e}\right)\phi_{,\eta} \nonumber \\
        & & \hspace{-40pt}
            -{4\over \sin_{\kappa}
            \left( \eta_{\rm o}-\eta_{\rm e}\right)}
            \int_{\eta_{\rm o}}^{\eta_{\rm e}}d\eta
            \, \sin_{\kappa}\left( \eta -\eta_{\rm
            e}\right)\phi_{,\eta } \nonumber \\
        & & \hspace{-40pt}
            -{4\over \sin_{\kappa}
            \left( \eta_{\rm o}-\eta_{\rm e}\right)}
            \int_{\eta_{\rm o}}^{\eta_{\rm e}}d\eta
            \, \cos_{\kappa}\left( \eta -\eta_{\rm e}\right)\phi
            \nonumber \\
        & & \hspace{-40pt}
            +{2\over \sin_{\kappa}
            \left( \eta_{\rm o}-\eta_{\rm e}\right)}
            \int_{\eta_{\rm o}}^{\eta_{\rm e}}d\eta
            \, \sin_{\kappa}\left( \eta -\eta_{\rm e}\right){\hat
            r}\gamma \left[ \phi_{,{\hat x}{\hat x}}+\phi_{,{\hat
            y}{\hat y}} \right]
  \label{eq:path2}
\end{eqnarray}
which is achieved by noting that the directional derivative along
$x^{(0)}$, $d\phi /d\eta = \phi_{,\eta} -\gamma \phi_{{\hat z}}$,
and performing an integration by parts.

While (\ref{eq:sol}) is, in principle, a complete description of the
angular-diameter distance in an FLRW spacetime perturbed by scalar
perturbations, it suffers from a number of defects that limit its
practical utility for observational cosmologists. Most seriously, it
exhibits explicit dependence on the unobservable point of emission
${\cal E}$ rather than on the observable redshift. Less seriously, it
does not explicitly depend on the usual cosmological parameters. We
proceed to remedy both defects.

Let ${\cal Q}$ at conformal time $\eta_{\rm q}$ and at spatial
position $(0, 0, {\hat r}(\eta_{\rm q}))$ on $x^{(0)}$ be an event
with the property that the redshift of a source at ${\cal Q}$
observed at ${\cal O}$ and computed in the background metric is
numerically the same as that of a source at ${\cal E}$ observed at
${\cal O}$ computed in the full perturbed spacetime. Since
$\eta_{\rm e}$ and $\eta_{\rm q}$ will differ only at first order,
and (\ref{eq:path2}) is valid only to this order, we can replace
$\eta_{\rm e}$ by $\eta_{\rm q}$ in (\ref{eq:path2}) without further
modification. The metric two-form factors are trickier, however.
Pyne \& Birkinshaw (1996) showed how to
calculate the coordinate displacements between ${\cal E}$ and
${\cal Q}$ suffered by a photon due to the presence of the
perturbations, $(\delta \eta , \delta {\hat x}^i)$. The components
of the displacement needed here are those corresponding to
conformal time and longitudinal coordinate ${\hat z}$. They are
\begin{eqnarray}
  \delta \eta     &=& {a_{\rm q}\over {\dot a}_{\rm q}}\left(
    v^{\hat z}_{\rm q}\gamma_{\rm e} -v_{\rm o}^{\hat z}\gamma_{\rm o}
    + \phi_{\rm q} -\phi_{\rm o}
    + k^{(1)0}_{\rm q}- k^{(1)0}_{\rm o}\right)
    \label{eq:disp1}
    \\
  \delta {\hat z} &=& -\gamma_{\rm q} \left( \delta \eta +I_S\right)
  \label{eq:disp2}
\end{eqnarray}
where we have used a dot to denote a conformal-time derivative. In
the above, the perturbation to the time-like component of the
photon wavevector
\begin{equation}
  k^{(1)0} = - 2\phi + 2\int_{\eta_{\rm o}}^{\eta} d\eta \,
             \phi_{,\eta}
\end{equation}
and
\begin{equation}
  I_S = 2\int_{\eta_{\rm o}}^{\eta_{\rm q}} d\eta \, \phi
\end{equation}
is the usual expression for the Shapiro delay, both expressions
being taken over the background path $x^{(0)}$.\footnote{In the
notation of Pyne \& Birkinshaw (1996), $\delta \eta =\delta
\lambda_{\rm q} +x^{(1)0}(\lambda_{\rm e})$ and $\delta x^i =
x^{(0)i}(\lambda_{\rm e})-x^{(0)i}(\lambda_{\rm q})
+x^{(1)i}(\lambda_{\rm e})$. By Taylor expansion, $\delta x^i
\approx k^{(0)i}_{\rm q}\delta\lambda_{\rm q}
+x^{(1)i}(\lambda_{\rm e})$. An integration by parts on the
expression for $x^{(1)i}(\lambda_{\rm e})$ given in Pyne \&
Birkinshaw then produces (\ref{eq:disp1},\ref{eq:disp2}).} We point
out that the 
displacement $\delta \eta$ is easily interpreted as the sum of
three effects: a differential doppler shift, a differential
gravitational redshift, and what is commonly termed the
``integrated Sachs-Wolfe effect" in studies of the CMBR.

The expressions (\ref{eq:disp1},\ref{eq:disp2}) allow us to Taylor
expand the metric 
two-form factors about their values at ${\cal Q}$ with the
result\footnote{We note that the transverse derivatives in the
Taylor expansion do not contribute once they are evaluated at
${\cal Q}$.}
\begin{eqnarray}
  {a^2_{\rm e}{\hat r}^2_{\rm e}\over \gamma^2_{\rm e}}
    &\approx &
      {a^2_{\rm q}{\hat r}^2_{\rm q}\over \gamma^2_{\rm q}}+
      \left( {a^2{\hat r}^2\over \gamma^2}\right)_{,\eta}\delta \eta
      +\left( {a^2{\hat r}^2\over \gamma^2}\right)_{,{\hat z}}\delta
      {\hat z} \phantom{\Biggl( \Biggr)} \nonumber \\
    & & \qquad \null \nonumber \\
    &=&
      {a^2_{\rm q}{\hat r}^2_{\rm q}\over \gamma^2_{\rm q}}
      +2{{\dot a}_{\rm q}\over a_{\rm q} }
        \left( {a^2_{\rm q}{\hat r}^2_{\rm q}\over \gamma^2_{\rm
        q}}\right)
        \delta \eta \nonumber \\
    & & \qquad \qquad 
      +2\left( {a^2_{\rm q}{\hat r}^2_{\rm q}\over
      \gamma^2_{\rm q}}\right){ \cos_\kappa (\eta_{\rm o} - \eta_{\rm
      q})\over {\hat r}_{\rm q}}\delta {\hat z}
      \phantom{\Biggl( \Biggr)}
  \label{eq:expand}
\end{eqnarray}
Recognizing that along the background path ${\hat r}/\gamma =
\sin_\kappa (\eta_{\rm o} -\eta )$ we see
\begin{eqnarray}
  {a^2_{\rm e}{\hat r}^2_{\rm e}\over \gamma^2_{\rm e}} 
    &\approx&
      {a^2_{\rm q}{\hat r}^2_{\rm q}\over \gamma^2_{\rm q}} \,
        \bigl[ 1+2{{\dot a}_{\rm q}\over a_{\rm q} }
             \delta \eta -2\cot_\kappa (\eta_{\rm o} - \eta_{\rm q} )
             \delta \eta \nonumber \\
    &       & \qquad -2\cot_\kappa (\eta_{\rm o} - \eta_{\rm q} )
              I_S \bigr] \quad .
  \label{eq:twoform}
\end{eqnarray}
Since $(1+z)^2 a_{\rm q}{\hat r}_{\rm q}/\gamma_{\rm q} = d_{\rm
L}^{(0)}$, the luminosity distance of the background cosmological
model, equations (\ref{eq:sol}), (\ref{eq:path2}) and
(\ref{eq:twoform}) lead to
\begin{equation}
  d_{\rm L} = d_{\rm L}^{(0)} (1+\delta )
  \label{eq:deltaexpr}
\end{equation}
with
\begin{eqnarray}
  \delta &=& v_{\rm o}^{\hat z} + {{\dot a}_{\rm q}\over a_{\rm q} }
             \delta \eta - \cot_\kappa (\eta_{\rm o} -\eta_{\rm q} )
             \delta \eta \nonumber \\
         & & \hspace{-24pt}
             - \cot_\kappa (\eta_{\rm o} -\eta_{\rm q} )
             I_S - \phi_{\rm q} -2 \phi_{\rm o}
             \nonumber \\
         & & \hspace{-24pt}
             +{2 \over \sin_\kappa \left( \eta_{\rm o} -
             \eta_{\rm q}\right)} \int_{\eta_{\rm o}}^{\eta_{\rm
             q}} d\eta \, \left( \eta -\eta_{\rm q}
             \right)\phi_{,\eta}
             \nonumber \\
         & & \hspace{-24pt}
             -{2\over \sin_{\kappa}
             \left( \eta_{\rm o} - \eta_{\rm q} \right)}
             \int_{\eta_{\rm o}}^{\eta_{\rm q} }d\eta
             \, \sin_{\kappa}\left( \eta -\eta_{\rm q}\right)
             \phi_{,\eta}
             \nonumber \\
         & & \hspace{-24pt}
             -{2\over \sin_{\kappa}
             \left( \eta_{\rm o} - \eta_{\rm q} \right)}
             \int_{\eta_{\rm o}}^{\eta_{\rm q}}d\eta
             \, \cos_{\kappa}\left( \eta -\eta_{\rm q}\right)\phi
             \nonumber \\
         & & \hspace{-24pt}
             +{1\over \sin_{\kappa}
             \left( \eta_{\rm o}-\eta_{\rm q} \right)}
             \int_{\eta_{\rm o}}^{\eta_{\rm q}}d\eta
             \, \sin_{\kappa}\left( \eta -\eta_{\rm q}\right){\hat
             r}\gamma \left[ \phi_{,{\hat x}{\hat x}}+\phi_{,{\hat
             y}{\hat y}} \right]
  \label{eq:delta1}
\end{eqnarray}

We now show how each term on the right-hand side of the above
equation may be expressed in terms of observable quantities and
standard cosmological parameters. First we change to the more
familiar comoving coordinates $(\eta, x, y, z)$ in which the
background metric assumes the form
\begin{equation}
  a^2(\eta ) \left( d\eta +{dr^2\over \sqrt{1-\kappa r^2}}
    +r^2 d\theta^2 + r^2\sin\theta d\phi^2 \right) \quad .
\end{equation}
Here $r=\sqrt{x^2+y^2+z^2}$ is the comoving radial coordinate distance
from the origin.\footnote{Some authors reserve this term for $a_{\rm
o} r$ which is a proper distance on the $\eta=\eta_{\rm o}$
hypersurface. $r$ here is a dimensionless coordinate distance.} This
change is effected by the transformation $x^i
={\hat x}^j / \gamma $. A number of useful results can now be
derived. First,  along $x^{(0)}$, $r=\sin_\kappa( \eta_{\rm o} -\eta
)$ and, in particular, $\sin_\kappa (\eta_{\rm o} -\eta_{\rm q}
)=r_{\rm q}$, the comoving coordinate distance to $\cal Q$. From
this follows $\cos_\kappa (\eta_{\rm o} -\eta ) =\sqrt{1-\kappa r^2}$
and the  change in measure, $dr =-\cos_\kappa (\eta_{\rm o} -\eta
)d\eta$.

On $x^{(0)}$, the Jacobian of the coordinate change may be shown to
take the value
\begin{eqnarray}
  {\partial x^i \over \partial {\hat x}^j}
    &=&
      \pmatrix{ {1 \over \gamma} & 0 & 0 \cr
                0 & {1 \over \gamma} & 0 \cr
                0 & 0 & {1 \over \gamma} - {\kappa \over 2}
                  {{\hat z}^2\over \gamma^2}}
  \label{eq:jacobian}
\end{eqnarray}
which secures, again on $x^{(0)}$,
\begin{eqnarray}
  \phi_{,xx} &=& \gamma^2\phi_{{\hat x}{\hat x}} \\
  \phi_{,yy} &=& \gamma^2\phi_{{\hat y}{\hat y}}
\end{eqnarray}

Finally, it will be convenient to employ the common formulae for
trigonometric functions of angular sums, in the forms
\begin{eqnarray}
  \sin_{\kappa}\left( \eta - \eta_{\rm q} \right)
    &=&
        \sin_{\kappa}\left( \eta - \eta_{\rm o}\right)
        \cos_{\kappa}\left( \eta_{\rm o} -\eta_{\rm q}\right) 
        \nonumber \\
    & & \qquad
        +\cos_{\kappa}\left( \eta -\eta_{\rm o}\right)
        \sin_{\kappa}\left( \eta_{\rm o} -\eta_{\rm q}\right) 
        \nonumber \\
    &=&
        -r\sqrt{1-\kappa r_{\rm q}^2}+r_{\rm q}\sqrt{1-\kappa r^2} \\
  \cos_{\kappa}\left( \eta -\eta_{\rm q}\right)
    &=&
        \cos_{\kappa}\left( \eta -\eta_{\rm o}\right)
        \cos_{\kappa}\left( \eta_{\rm o} -\eta_{\rm q}\right)
        \nonumber \\
    & & \qquad
       -\kappa \sin_{\kappa}\left( \eta -\eta_{\rm o}\right)
               \sin_{\kappa}\left( \eta_{\rm o} -\eta_{\rm q}\right)
       \nonumber \\
    &=&
        \sqrt{1-\kappa r_{\rm q}^2}\sqrt{1-\kappa r^2}-\kappa r r_{\rm q} 
        \\
  \eta -\eta_{\rm q}
    &=&
        -(\eta_{\rm o} -\eta ) +(\eta_{\rm o} -\eta_{\rm q} ) 
        \nonumber \\
    &=&
        -\sin_\kappa^{-1} r +\sin^{-1}_\kappa r_{\rm q} \quad .
\end{eqnarray}
Using these identities, we find
\begin{eqnarray}
  \delta
    &=&
        \left( v^z_{\rm q} - \phi_{\rm o} - 2 \phi_{\rm q}
        +I_{ISW} \right) 
        \nonumber \\
    & & \quad
        -r_{\rm q}^{-1} \sqrt{1-\kappa r_{\rm q}^2} \, \delta \eta
        -r_{\rm q}^{-1} \sqrt{1-\kappa r_{\rm q}^2} \, I_S 
        \nonumber \\
    & & \quad
        - 2 r_{\rm q}^{-1} \int_{0}^{r_{\rm q}}{dr \over \sqrt{1-\kappa
        r^2}}
        \left[ -\sin_\kappa^{-1} r +\sin^{-1}_\kappa r_{\rm q} \right]
        \phi_{,\eta} 
      \nonumber \\
    & & \quad
        + 2 r_{\rm q}^{-1}
        \int_{0}^{r_{\rm q}}{dr\over \sqrt{1-\kappa r^2}}
        \, \bigl[ -r\sqrt{1-\kappa r_{\rm q}^2}
        \nonumber \\
    & & \quad \quad 
                +r_{\rm q}\sqrt{1-\kappa r^2} \bigr] \phi_{,\eta} 
        \nonumber \\
    & & \quad
        + 2 r_{\rm q}^{-1}
        \int_{0}^{r_{\rm q}}{dr\over \sqrt{1-\kappa r^2}}
        \, \bigl[ \sqrt{1-\kappa r_{\rm q}^2}\sqrt{1-\kappa r^2}
        \nonumber \\
    & & \quad \quad
        -\kappa r r_{\rm q} \bigr] \phi
        \nonumber \\
    & & \quad
      - r_q^{-1}
      \int_{0}^{r_{\rm q}}{r \, dr\over \sqrt{1-\kappa r^2}}
      \, \bigl[ -r\sqrt{1-\kappa r_{\rm q}^2}
       \nonumber \\
    & & \quad \quad +r_{\rm q}\sqrt{1-\kappa
      r^2} \bigr] \left[ \phi_{,xx}+\phi_{,yy} \right]
  \label{eq:sol2}
\end{eqnarray}
where
\begin{equation}
  I_{ISW} = -2 \int_{0}^{r_{\rm q}} {dr \over \sqrt{1-\kappa r^2}}
            \, \phi_{,\eta}
  \label{eq:iisw}
\end{equation}
is the integrated Sachs-Wolfe effect,
\begin{equation}
  I_S = - 2 \int_{0}^{r_{\rm q}} {dr\over \sqrt{1-\kappa r^2}} \, \phi
  \label{eq:shapiro}
\end{equation}
is the Shapiro effect and
\begin{equation}
  \delta \eta ={a_{\rm q}\over {\dot a}_{\rm q}}\left( v^z_{\rm q}
  -v_{\rm o}^z + \phi_{\rm o} - \phi_{\rm q} + I_{ISW} \right) 
  \quad . 
\end{equation}

\begin{figure}
 \epsfxsize 8.45cm
 \epsfbox{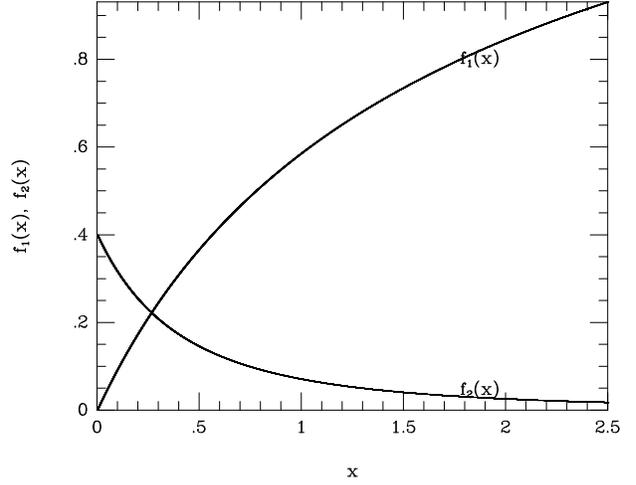}
 \epsfxsize 8.45cm
 \epsfbox{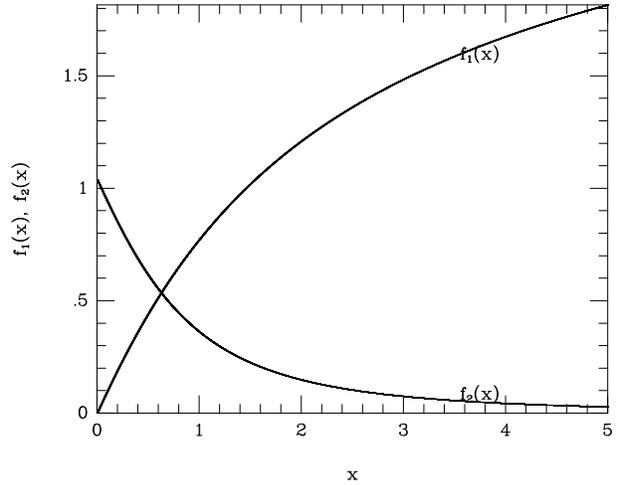}
 \caption{The functions $f_1(x)$ and $f_2(x)$ for
  $(\Omega_{\rm m}, \Omega_\Lambda)=(1.0,0.0)$ (top) and
  $(\Omega_{\rm m}, \Omega_\Lambda)=(0.3,0.7)$ (bottom).}
 \label{fig:f1f2}
\end{figure}

Because of the way in which we defined ${\cal Q}$, the observable
redshift of our source obeys the standard equation $1+z = a_{\rm
o}/a_{\rm q}$ and many of the standard cosmological equations apply
with their usual interpretation. In particular, the Hubble factor at
the location of a source of redshift $z$ is given by\footnote{It is
important to realize that the numerator here is a conformal time
derivative. This accounts for the unusual power in the denominator.}
\begin{equation}
  H = { {\dot a} \over a^2} = H_{\rm o} E(z)
  \label{eq:hubble}
\end{equation}
with
\begin{equation}
  E(z) =\sqrt{(1+z)^2 (1+z\Omega_{\rm m}) -z(2+z) \Omega_\Lambda } \quad .
\end{equation}
Additionally, the comoving proper distance to $\cal Q$ is
\begin{equation}
  a_{\rm o} r_{\rm q} = {1 \over H_{\rm o} \sqrt{ | \Omega_\kappa
  |}}\sin_\kappa \left( \sqrt{ | \Omega_\kappa |} f_1(z) \right)
  \label{eq:proper}
\end{equation}
where $f_1(z)$ is the integral (shown in Fig.~\ref{fig:f1f2})
\begin{equation}
  f_1(z) = \int_0^z {dl \over E(l)}
  \label{eq:f1def}
\end{equation}
and we have the usual expressions for the density parameters in
matter, vacuum energy, and curvature
\begin{equation}
  \Omega_{\rm m} = {8\pi \over 3 H_0^2} \rho_{\rm m0}
  \qquad
  \Omega_\Lambda = {\Lambda \over 3H_0^2}
  \qquad
  \Omega_\kappa  = - {\kappa \over a_{\rm o}^2 H_0^2}
\end{equation}
respectively. As usual, these constants are related by the identity
$\Omega_{\rm m} +\Omega_\Lambda +\Omega_\kappa = 1$ which has the
crucial consequence that for $\Omega_\kappa \neq 0$
\begin{eqnarray}
  a_{\rm o} H_{\rm o} &=& \sqrt{ |\Omega_\kappa^{-1} | }\nonumber \\
                      &=& \sqrt{ | (1 - \Omega_{\rm m}
                                      - \Omega_\Lambda )^{-1} | }
\end{eqnarray}
and hence $a_{\rm o}$ is expressible in terms of the usual
cosmological parameters. This is in contrast to the flat case
where there is no meaningful curvature scale available.

For the curved cases, then, both $r_{\rm q}$ and $a_{\rm o} r_{\rm
q}$ are functions of the standard cosmological parameters and the
observed source redshift and so (\ref{eq:sol2}) represents the
desired formula for the luminosity distance. Simply for clarity we
record, for these cases,
\begin{equation}
  r_{\rm q}=\sin_\kappa \left( \sqrt{ | \Omega_\kappa |} f_1(z) \right)
\end{equation}
and, valid for all cases, curved and flat,
\begin{equation}
  r_{\rm q}^{-1}\delta \eta ={ \sqrt{| \Omega_\kappa |} \, (1+z) \,
    \left( v^z_{\rm q} - v_{\rm o}^z +\phi_{\rm o} -\phi_{\rm q}
    +I_{ISW} \right) \over E(z) \sin_\kappa \left( \sqrt{ |
    \Omega_\kappa |} f_1(z) \right)}
\end{equation}
In the flat case our formula simplifies considerably, taking the form
\begin{eqnarray}
  \delta
    &=& \left( v^z_{\rm q} - \phi_{\rm o} - 2 \phi_{\rm q}
        + I_{ISW}\right)
        \nonumber \\
    & & \quad
        - r_{\rm q}^{-1}\delta \eta - 2 r_{\rm q} ^{-1}I_S
        \nonumber \\
    & & \quad
        - r_{\rm q}^{-1} \int_{0}^{r_{\rm q}} dr
        \, \left[ -r +r_{\rm q} \right] r\left[
            \phi_{,xx}+\phi_{,yy} \right]
  \label{eq:sol3}
\end{eqnarray}
In contrast to the curved cases, in the flat case it is not
possible to express $r_{\rm q}$ in terms of standard cosmological
parameters and source redshift alone. As a result, it looks at
first as if the terms in (\ref{eq:sol3}) involving integrations will
prove problematic to our program. In fact, however, this
difficulty is easily overcome by the change of variables $r
\mapsto r_{\rm q}w$ and $k=|{\bf k}|\mapsto v/r_{\rm q}$ performed
after Fourier decomposition of $\phi$ with mode-variable $\bf k$.
This procedure is illustrated in the examples below.

\section{Luminosity distance scatter from weak lensing}
\label{sec:lensing}

The last term in (\ref{eq:sol3}), and its antecedent in (\ref{eq:sol2}),
are precisely the standard cosmological weak lensing terms usually
analyzed in isolation. In this section we illustrate how
these terms can be expressed as functions of the standard
cosmological parameters, and work out the contribution of the last
term in (\ref{eq:sol3}) numerically for a simple $\Omega_{\rm m}=1$
flat cosmology, and for the currently popular flat $(\Omega_{\rm
m}, \Omega_\Lambda)=(0.3,0.7)$ cosmology. The identification and
analysis of the weak lensing term using a perturbed geodesic
equation is not novel (Bernadeau, Van Waerbeke \& Mellier 1997; Kaiser
1998). Its emergence as the dominant term in a physically sensible and 
complete description of the luminosity distance is, however, and
we include its analysis here for completeness as it is somewhat
tricky to cast in the proper form.

In order to evaluate
\begin{equation}
  \delta_{\rm lens}=\int_{0}^{r_{\rm q}}dr
    \, \left( {r\over r_q}-1\right)r
    \left[\phi_{,xx}+ \phi_{,yy}\right]
\end{equation}
in either cosmological model it is necessary to specify the potential
perturbation, $\phi$. $\phi$ satisfies the perturbed Einstein's
equations in the longitudinal gauge as given, for example, as
equations (5.17)-(5.19) in Mukhanov et al. (1992) for $\Omega_\Lambda
=0$ cosmologies. For the flat $\Omega_{\rm m}=1$ cosmology during an
epoch of matter-domination, and when the hydrodynamical
perturbations described by $\phi$ are pressureless, $\phi$ is
independent of conformal time, taking the form $\phi ( {\bf x}
,\eta)=\phi( {\bf x} )$.\footnote{It is possible to express the
Laplacian of $\phi ({\bf x})$ in terms of the fractional perturbation
to the hydrodynamical energy density on a surface of constant
conformal time via what is usually termed a ``cosmological Poisson
equation" but this is not necessary here.} For models with vacuum
energy, $\phi$ varies with conformal time according to
$\phi ( {\bf x} ,\eta)=s(\eta )\phi( {\bf x} )$ with $s(\eta)$
obtained from
\begin{equation}
  s(z) = (1+z) \, E(z) \, {f_2(z) \over f_2(0)}
  \label{eq:s}
\end{equation}
where $f_2(z)$ is the integral
\begin{equation}
  f_2(z) = \int_z^\infty {(1 + l) \, dl \over E(l)^3} \quad ,
\end{equation}
(see Fig.~\ref{fig:f1f2}) and $z(\eta)$ is given through inversion of
\begin{equation}
  f_1(z) =a_{\rm o} H_0 (\eta_{\rm o} -\eta) \quad .
  \label{eq:z}
\end{equation}
The factor $f_2(0)$ in (\ref{eq:s}) ensures the conventional
normalization for the $\Omega_{\rm m}=1$ case.

While the temporal variation of the potential is deterministic in
the models considered, the spatial variation is unconstrained. It
is common to imagine the actual $\phi ({\bf x})$ as one
realization drawn from a Gaussian random field. In this picture,
the random nature of $\phi$ is usually expressed by specifying the
probability distributions of its Fourier components,
\begin{equation}
  \phi ( {\bf x} , \eta ) =s(\eta )\int_{V}  h ( {\bf k} ) e^{i {\bf
     k}\cdot {\bf x}} {d^3{\bf k}\over (2\pi)^3}
\end{equation}
with
\begin{eqnarray}
  \langle h({\bf k}) h^* ({\bf k}')\rangle &=&(2\pi)^3 \delta ({\bf
            k}- {\bf k}' ) P(k) \\
  \langle h({\bf k}) h ({\bf k}')\rangle   &=&
           (2\pi)^3 \delta ({\bf k}+ {\bf k}' ) P(k) \\
  \langle h^*({\bf k}) h^* ({\bf k}')\rangle &=& (2\pi)^3 \delta ({\bf
    k}+ {\bf k}' ) P(k)
\end{eqnarray}
where the last two correlators follow from the first because of the
requirement that $\phi$ be real.

Now we are ready to compute the RMS fractional deviation due to
lensing. It is easiest to take the various correlators into account by
writing $\phi$ in the form
\begin{eqnarray}
  \phi ( {\bf x} , \eta)
    &=&
      {1\over 2} \phi( {\bf x} , \eta )
      +{1\over 2} \phi( {\bf x} , \eta )^*
      \nonumber \\
    &=&
      {s(\eta )\over 2}\int_{V}  h ( {\bf k} ) e^{i {\bf k}\cdot {\bf
      x}} {d^3{\bf k}\over (2\pi)^3}
      \nonumber \\
    & & \quad +{s(\eta )\over 2}\int_{V}
      h^* ( {\bf k} ) e^{-i {\bf k}\cdot {\bf x}} {d^3{\bf
      k}\over (2\pi)^3}
\end{eqnarray}
whence
\begin{eqnarray}
  \langle \delta_{\rm lens}^2 \rangle
    &=& {1\over  (2\pi )^3} \int d^3{\bf k} P(k)\left[
        k_x^2 +k_y^2\right]^2
        \nonumber \\
    & & \quad \times \Bigg| \int_{0}^{r_q}dr
        \, \left( {r\over r_q}-1\right)rs(r)
        e^{i k_zr} \Bigg|^2
  \label{eq:lambda}
\end{eqnarray}
with $s(r)$ given by the compositions of (\ref{eq:s}) and (\ref{eq:z})
above with the identity, true along $x^{(0)}$ where we need it,
$(\eta_{\rm o} -\eta)=r$.

\subsection{Lensing effects from a simple power spectrum}
\label{sec:simplelens}

We must know the potential power spectrum $P(k)$ to perform the
integral in (\ref{eq:lambda}). For the purposes of illustration, we
follow Seljak (1994) and assume a broken power-law form
\begin{equation}
  P(k) =\cases{ A k^{-3}       & for $k\le k_0$ \cr
                A k_0^4 k^{-7} & for $k\ge k_0$ \cr
              }
  \label{eq:spectrum}
\end{equation}
with $k_0$, the turnover wavenumber, corresponding to a physical scale
of about $10$ Mpc. The constant $A$ is set by the CMBR quadrupole of
$Q_2^2=(6\times 10^{-6})^2$ by
\begin{equation}
  Q_2^2={20\pi K_2^2 \over 9}\int_0^\infty k^2 P(k) j_2^2 (2k/H_0)\,
        dk
  \label{eq:q2}
\end{equation}
(Bond \& Efstathiou 1987). Here $j_2(x)$ is the spherical Bessel
function of order 2 and $K_2^2$ is the amplification coefficient of
Kofman \& Starobinskii (1985) which accounts for
the time variation of the potential. For the $\Omega_{\rm m}=1$ case,
$K_2=1$ and for the $(\Omega_{\rm m},\Omega_\Lambda)=(0.3,0.7)$ case
$K_2=1.19$. The normalization for the power spectrum then becomes
\begin{equation}
  A = \cases{6.2\times 10^{-11} & if $(\Omega_{\rm m},\Omega_\Lambda)
                                  =(1,0)$ \cr
             4.4\times 10^{-11} & if $(\Omega_{\rm m},\Omega_\Lambda)
                                  =(0.3,0.7)$.
            }
\end{equation}

We simplify the notation by generalizing (\ref{eq:spectrum}) to
\begin{equation}
  P(k) = A k_0^{n-3} k^{-n}
\end{equation}
which can be used both above and below the turnover wavenumber, and
change variables in the radial integration to $w = r/r_{\rm q}$ and in
the $|{\bf k}|$ integration to $v=k r_{\rm q}$. Inserting this into
(\ref{eq:lambda}), we obtain
\begin{eqnarray}
  \langle \delta^2_{\rm lens} \rangle
    &=&
      {A (k_0 r_{\rm q})^{n-3}
      \over (2\pi)^2} \int dv \, v^{6-n} \int_{-1}^1
      du (1-u^2)^2 \\
    & & \quad
      \times \Bigg| \int_{0}^1 dw \, w \, (w-1) \, s(w,z) \, e^{i uvw}
      \, \Bigg|^2
\end{eqnarray}
for the contributions to $\langle \delta^2_{\rm lens} \rangle$ at
large and small $k$. In this equation, $s(w,z)$ given by the
composition of $s({\tilde z})$, given in equation (\ref{eq:s}), with
${\tilde z}(w,z)$, obtained by solving
\begin{equation}
  w f_1(z) = f_1({\tilde z}) \quad .
\end{equation}
It may be worth noting here that $\tilde z$ is simply an intermediate
variable that will disappear from the final result whereas $z$ is the
observable redshift of the source. 

The source redshift and the cosmological parameters $\Omega_{\rm
m}$ and $\Omega_\Lambda$ can be seen to enter into $\langle
\delta_{\rm lens}^2 \rangle$ in the same manner: that is, via
$s(w,z)$ and the product $v_0 \equiv k_0 r_{\rm q}$ which is equal
to $2\pi$ times the comoving proper distance to $\cal Q$ in 
units of the turnover wavelength
\begin{eqnarray}
  v_0 \equiv k_0 r_{\rm q}
    &=&
      {2\pi \over \lambda_0 H_0} \int_0^z {dl \over E(l)} \cr
    &=&
      1884 \, h^{-1} \, f_1(z) \quad .
\end{eqnarray}

If we use the values of $n$ from (\ref{eq:spectrum}), then the total
luminosity distance scatter $\langle \delta^2_{\rm lens} \rangle$ is
given by the explicit form
\begin{eqnarray}
 \langle \delta^2_{\rm lens} \rangle
    &=& {A \over 4 \pi^2} \, \int_0^{v_0} dv \, v^3 \,
             \int_{-1}^{+1} du \, (1 - u^2)^2 \nonumber \\
    & & \quad \times \int_0^1 dw \, w (w-1) \int_0^1 dw^\prime \,
                            w^\prime (w^\prime-1) \nonumber \\
    & & \quad \times s(w,z) \, s(w^\prime,z) \,
              \cos \left( uv(w-w^\prime) \right) \nonumber \\
    & & + {A \over 4 \pi^2} \, v_0^4 \,
             \int_{v_0}^{\infty} {dv \over v} \,
             \int_{-1}^{+1} du \, (1 - u^2)^2 \, \nonumber \\
    & & \quad \times \int_0^1 dw \, w (w-1) \int_0^1 dw^\prime \,
                            w^\prime (w^\prime-1) \nonumber \\
    & & \quad \times s(w,z) \, s(w^\prime,z) \,
              \cos \left( uv(w-w^\prime) \right) \quad .
 \label{eq:deltadef}
\end{eqnarray}

\begin{figure}
 \epsfxsize 8.45cm
 \epsfbox{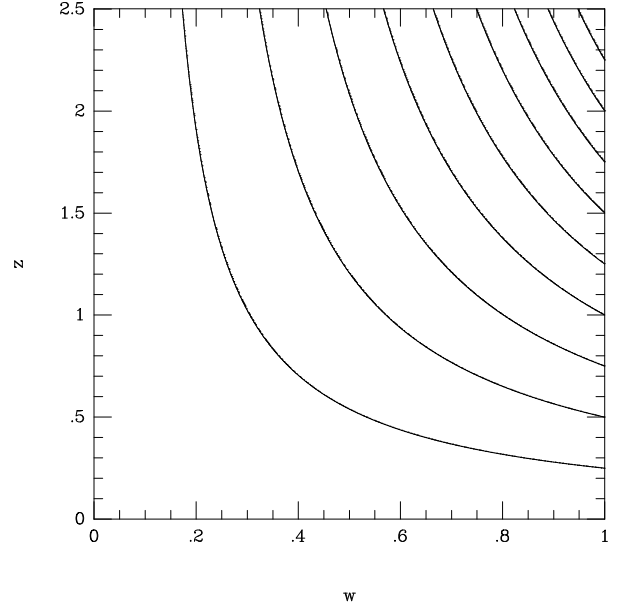}
 \caption{The function $\tilde{z}(z,w)$ for
  $(\Omega_{\rm m}, \Omega_\Lambda)=(0.3,0.7)$. Contours are drawn at
 $\tilde{z} = 0.5$, $1.0$, $\ldots$, $4.5$.}
 \label{fig:ztilde}
\end{figure}

\begin{figure}
 \epsfxsize 8.45cm
 \epsfbox{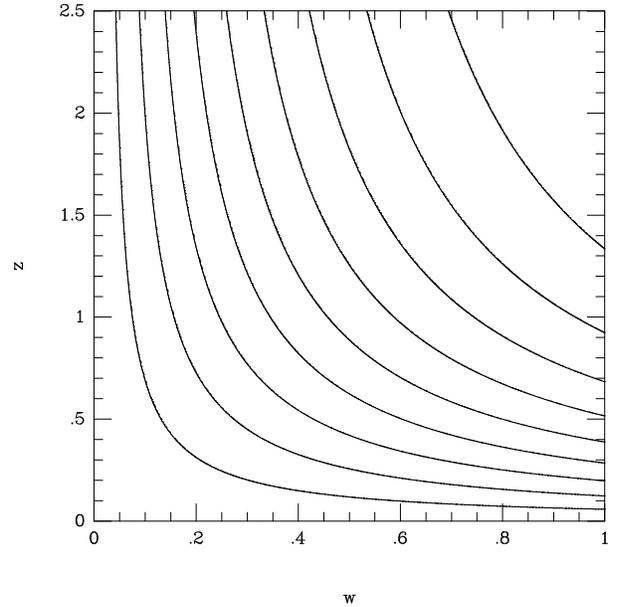}
 \caption{The function $s(z,w)$ for
  $(\Omega_{\rm m}, \Omega_\Lambda)=(0.3,0.7)$. For
  $(\Omega_{\rm m}, \Omega_\Lambda)=(1.0,0.0)$, $s = {2 \over 5}$ for
  all redshift. Contours are drawn at $s = 1.07,\ 1.10,\dots ,1.29$.}
 \label{fig:sfuncs}
\end{figure}

While the calculation of the functions $f_1$ and $f_2$
(Fig~\ref{fig:f1f2}), $\tilde{z}$ (Fig.~\ref{fig:ztilde}), and $s$ 
(Fig.~\ref{fig:sfuncs}) present no difficulties, the four-dimensional
integrals in (\ref{eq:deltadef}) have some unpleasant characteristics,
which can be brought out by further analysis.

It is convenient to change the order of integration in
(\ref{eq:deltadef}), starting by performing the $u$-integral, which
can be done analytically. Define
\begin{eqnarray}
 g(x) &\equiv& \int_{-1}^{+1} du \, \left( 1 - u^2 \right)^2 \,
               \cos \left( u x \right)
               \nonumber \\
      &=     & -{48 \over x^4}\cos x + {48 \over x^5} \sin x
               -{16 \over x^3} \sin x \quad .
\end{eqnarray}
Then integral (\ref{eq:deltadef}) can be rewritten
\begin{eqnarray}
 \langle \delta^2_{\rm lens} \rangle
  &=& {A \over 4 \pi^2} \, v_0^4 \,
      \int_0^1 dw        \, w        (w       -1) \, s(w       ,z) \,
      \nonumber \\
  & &\quad \times
      \int_0^1 dw^\prime \, w^\prime (w^\prime-1) \, s(w^\prime,z) \,
      t(q)
 \label{eq:deltared}
\end{eqnarray}
where the function
\begin{eqnarray}
 t(q) &=& \int_0^1 dy \, y^3 \, g(yq) + \int_1^\infty {dy \over y} \,
          g(yq) \nonumber \\
      &=& \int_0^1 dy \, \left( y^3 \, g\left(yq\right)
          + {1 \over y} \, g\left({q \over y}\right) \right)
 \label{eq:tqdef}
\end{eqnarray}
and $q = v_0 \left( w - w^\prime \right)$. $t(q)$ is the key
function in what follows, but once calculated, the same $t(q)$
function can be used for all flat cosmologies, since the
cosmological behaviour is concealed in the scaling quantity $v_0$
and in the $(w,w^\prime)$ integrations via the $s(w,z)$ function.
Fortunately, the integral in (\ref{eq:tqdef}) can be performed, to
yield
\begin{eqnarray}
 t(q) &=& {16 \over 15} \, q^{-5} \, \bigl(
                           -36     \sin q
                           + 6 q   \cos q
                           +30 q
                           - 2 q^2 \sin q \nonumber \\
      & & \quad
                           -   q^3 \cos q
                           +   q^4 \sin q
                           -   q^5 {\rm Ci}(q) \bigr)
 \label{eq:tq}
\end{eqnarray}
where the reader is to understand $|q|$ for $q$ (formally,
the equation is slightly different for negative $q$, since $t(q)$
is even in $q$ but ${\rm Ci}(q)$ is defined with a cut along
the negative $q$ axis). The cosine integral function ${\rm Ci}(q)$
appearing in (\ref{eq:tq}) causes the major difficulty in evaluating
the integral (\ref{eq:deltared}) because as $q \rightarrow 0$,
\begin{equation}
 t(q) \rightarrow {16 \over 15} \left( {107 \over 60} - \gamma -\ln q
                   + {1 \over 42}q^2
                   - {1 \over 4032}q^4 + O(q^6) \right)
 \label{eq:tqsmall}
\end{equation}
where $\gamma = 0.577\ldots$ is Euler's constant. This exposes the
logarithmic singularity at $q = 0$. By contrast, ${\rm Ci}(q)$ is
well-behaved as $q \rightarrow \infty$, with $| {\rm Ci}(q) |
\propto q^{-1}$, so that in this limit
\begin{equation}
 t(q) \rightarrow 32 \, q^{-4} \, \left( 1 - {2 \over q} \sin q + {4
 \over q^2} \cos q +O\left( q^{-3}\right) \right) \quad .
 \label{eq:tqlarge}
\end{equation}

\begin{figure}
 \epsfxsize 8.45cm
 \epsfbox{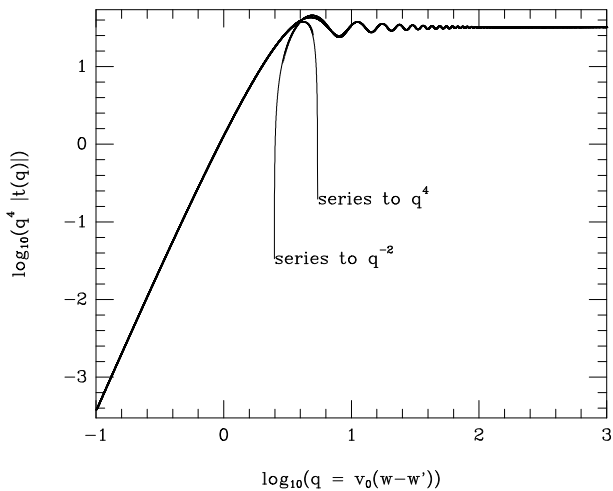}
 \caption{The function $q^4 t(q)$ for
  $(\Omega_{\rm m}, \Omega_\Lambda)=(0.3,0.7)$ and a range of $q$
  appropriate to the integral for redshift $z = 2.5$. Note the large
  range of $t(q)$, the singularity as $q \rightarrow 0$, and the
  oscillations in $t(q)$ at moderate and large $q$. This weighted
  $q^4 t(q)$ has a maximum at $q = 4.77$: $t(q)$ diverges as
  $-\ln q$ at small $q$.}
 \label{fig:tq}
\end{figure}

The form of $t(q)$ is shown in Fig.~\ref{fig:tq}: note that the
asymptotic forms (\ref{eq:tqsmall},\ref{eq:tqlarge}) are good
descriptions of the behaviour of $t(q)$, but that the logarithmic
divergence at $q \rightarrow 0$ makes the integration in
(\ref{eq:deltared}) improper, with implications for the numerical
scheme adopted.

\subsection{Numerical results}
\label{sec:numerical}

If we use (\ref{eq:spectrum}) for the potential power spectrum, then
the calculation of the integral for $\langle \delta^2_{\rm lens}
\rangle$ becomes a problem of covering the $(w,w^\prime)$ plane
adequately at $w \approx w^\prime$ so that the logarithmic singularity
and the oscillations of $t(q)$ for moderate $w-w^\prime$ are well
sampled. An efficient solution is to alter the integration variables
$(w,w^\prime)$ to $(w_{\rm a},w_{\rm b})$, where
\begin{eqnarray}
 w_{\rm a} &=& {1 \over \sqrt{2}} \left(w - w^\prime \right) \\
 w_{\rm b} &=& {1 \over \sqrt{2}} \left( 1 - w - w^\prime \right)
             \quad .
\end{eqnarray}
Since $q = v_0 \, w_{\rm a}$, and is independent of $w_b$, the
integration now needs to be done carefully only in the $w_{\rm a}$
direction, while the change in the integrand in (\ref{eq:deltared}) in
the $w_{\rm b}$ direction is slow. Furthermore, the symmetry
$(w,w^\prime) \rightarrow (w^\prime,w)$ implies that only the
triangle
\begin{eqnarray}
 w_{\rm a} &\in& 0,{1 \over \sqrt{2}} \\
 w_{\rm b} &\in& - \left({1 \over \sqrt{2}} - w_{\rm a}\right) ,
                   \left({1 \over \sqrt{2}} - w_{\rm a}\right)
\end{eqnarray}
needs to be included in the integral since the $w_{\rm a}<0$ triangle
gives the same contribution. For the case $(\Omega_{\rm
m},\Omega_\Lambda) = (1,0)$, a further simplification is possible,
and the smaller triangle
\begin{eqnarray}
 w_{\rm a} &\in& 0,{1 \over \sqrt{2}} \\
 w_{\rm b} &\in& 0, \left({1 \over \sqrt{2}} - w_{\rm a}\right)
\end{eqnarray}
is all that is needed.

\begin{figure}
 \epsfxsize 8.45cm
 \epsfbox{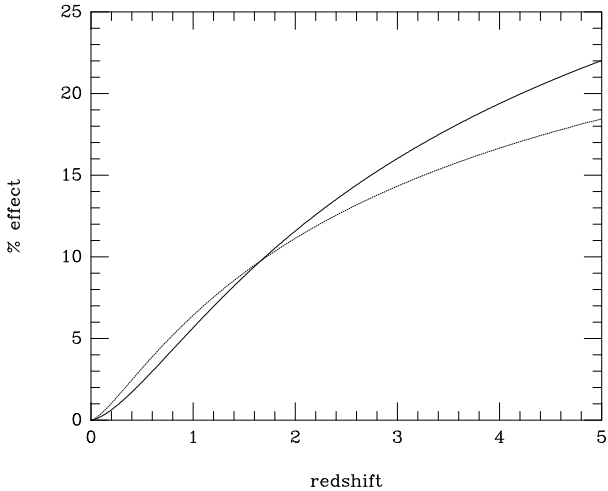}
 \caption{The value of
  $\sqrt{\langle \delta^2_{\rm lens} \rangle}$, expressed as a
  percentage change, as a function of redshift to $z = 2.5$ for
  $(\Omega_{\rm m}, \Omega_\Lambda)=(1,0)$ (dashed line) and
  $(\Omega_{\rm m}, \Omega_\Lambda)=(0.3,0.7)$ (solid line).}
 \label{fig:delta}
\end{figure}

The dashed line on Fig.~\ref{fig:delta} shows the result obtained
for $(\Omega_{\rm m},\Omega_\Lambda)=(1,0)$. The effect rises
approximately as
\begin{equation}
  \sqrt{\langle \delta^2_{\rm lens} \rangle} = 12 \times 10^{-2} \,
  z^{3 \over 2}
\end{equation}
at small $z$ (with $h = 0.72$), and reaches about 6\% at $z =
1$, and more than 10\% at $z = 2$. The solid line on
Fig.~\ref{fig:delta} shows the result for $(\Omega_{\rm m},
\Omega_\Lambda) = (0.3,0.7)$. The curve has a similar shape to that
for the Einstein-deSitter case, but larger 
lensing effects are seen at $z > 1.7$. At $z < 0.4$ this curve can be
represented to an accuracy of 4\% by 
\begin{equation}
  \sqrt{\langle \delta^2_{\rm lens} \rangle} = 7.5 \times 10^{-2} \,
  z^{3 \over 2} \quad .
\end{equation}
The results for both cosmological models show the $z^{3 \over 2}$
dependence that one would have expected on the basis of
(\ref{eq:deltadef}) alone. Numerical results for both cases (accurate
to better than 1\%) are given in Table~\ref{tab:results}.

\begin{table}
 \caption{Results for $\sqrt{\langle \delta^2_{\rm lens} \rangle}$ for
  the two cosmologies}
 \label{tab:results}
 \begin{tabular}{@{}c|cc}
  Redshift, $z$ & \multicolumn{2}{c}{$\sqrt{\langle \delta^2_{\rm
  lens} \rangle}$ (\%) for $(\Omega_{\rm m},\Omega_\Lambda)=$} \\
        & $(1.0,0.0)$      & $(0.3,0.7)$ \\ 
   0.01 &  0.014           &  0.007           \\
   0.02 &  0.040           &  0.021           \\
   0.03 &  0.073           &  0.038           \\
   0.04 &  0.11\phantom{0} &  0.059           \\
   0.05 &  0.15\phantom{0} &  0.082           \\
   0.06 &  0.20\phantom{0} &  0.11\phantom{0} \\
   0.07 &  0.25\phantom{0} &  0.14\phantom{0} \\
   0.08 &  0.30\phantom{0} &  0.17\phantom{0} \\
   0.09 &  0.35\phantom{0} &  0.20\phantom{0} \\
   0.10 &  0.41\phantom{0} &  0.23\phantom{0} \\
   0.20 &  1.05\phantom{0} &  0.64\phantom{0} \\
   0.30 &  1.77\phantom{0} &  1.16\phantom{0} \\
   0.40 &  2.50\phantom{0} &  1.74\phantom{0} \\
   0.50 &  3.22\phantom{0} &  2.37\phantom{0} \\
   0.60 &  3.93\phantom{0} &  3.02\phantom{0} \\
   0.70 &  4.61\phantom{0} &  3.70\phantom{0} \\
   0.80 &  5.26\phantom{0} &  4.37\phantom{0} \\
   0.90 &  5.89\phantom{0} &  5.05\phantom{0} \\
   1.00 &  6.49\phantom{0} &  5.73\phantom{0} \\
   1.20 &  7.62\phantom{0} &  7.04\phantom{0} \\
   1.40 &  8.64\phantom{0} &  8.31\phantom{0} \\
   1.60 &  9.59\phantom{0} &  9.51\phantom{0} \\
   1.80 & 10.5\phantom{00} & 10.6\phantom{00} \\
   2.00 & 11.3\phantom{00} & 11.7\phantom{00} \\
   2.50 & 13.0\phantom{00} & 14.0\phantom{00} \\
   3.00 & 14.5\phantom{00} & 16.2\phantom{00} \\
   3.50 & 15.7\phantom{00} & 18.0\phantom{00} \\
   4.00 & 16.8\phantom{00} & 19.6\phantom{00} \\
   4.50 & 17.8\phantom{00} & 21.0\phantom{00} \\
   5.00 & 18.6\phantom{00} & 22.3\phantom{00} \\
 \end{tabular}

\end{table}

\section{Summary}
\label{sec:summary}

We have presented a complete formula (\ref{eq:sol2}) for the
luminosity distance in linearly perturbed FLRW spacetimes. The
simpler form (\ref{eq:sol3}) is appropriate in flat spacetimes. These
results give the first explicit presentations of the various
gravitational effects which modulate the unperturbed 
luminosity distance in terms that can be related to observable
quantities, as shown explicitly for $(\Omega_{\rm m},\Omega_\Lambda) =
(1,0)$ and $(0.3,0.7)$ cosmologies in Section~\ref{sec:simplelens}.

The cosmological weak lensing term from (\ref{eq:sol3}) leads to a
fractional scatter in the luminosity distance, $\langle \delta_{\rm
lens}^2 \rangle$, which can be significant as shown in 
Table~\ref{tab:results} and Fig.~\ref{fig:delta}. Clearly this effect
is appreciable for quasars and high-redshift galaxies, especially as
such objects are now being seen to $z > 6$. The principal assumptions
required to derive these results are that the form (\ref{eq:spectrum})
is a good description of the potential power spectrum, and that the
time evolution is accurately described by $s(\eta)$.

At $z = 1$, the value of $\sqrt{\langle \delta^2_{\rm lens} \rangle}$
is roughly $0.06$. Objects of fixed absolute magnitude would therefore
show an apparent magnitude scatter of about 0.12~mag from lensing
alone. Since the scatter of SN~Ia absolute magnitudes (after
correction to a common light curve) is about 0.17~mag (Perlmutter et
al.~1999), it is clear that lensing makes a significant, and
increasingly important, contribution to the scatter of the SN~Ia
Hubble diagram as it is extended to redshifts $> 1$. At the
redshift limit of the SNAP mission ($z \sim 1.7$; Perlmutter et
al.~2003), the lensing-induced scatter of supernova apparent
magnitudes rises about $0.2$~mag, and becomes a dominant contribution
to the intrinsic noise in the Hubble diagram. The associated Malmquist
bias may also become important.

While the usual cosmological weak lensing term is dominant, other
terms are present and can be expected to affect the correlations
between luminosity distance corrections and other physical quantities
such as the integrated Sachs-Wolfe effect. The approach we have used,
direct integration of the null geodesic equation, can also be used to
examine the effects of vector and tensor perturbations, and can be
extended, with somewhat more difficulty, to higher orders using the
results of Pyne \& Carroll (1996).

\section{Acknowledgements} T.P. would like to thank Alex Barnett
for substantial help with numerical studies that informed an early
draft of the current paper.

\end{document}